\documentclass[]{spie}  

\usepackage{amsmath,amsfonts,amssymb}
\usepackage{graphicx}
\usepackage{float}
\usepackage[colorlinks=true, allcolors=blue]{hyperref}
\usepackage{lscape}
\title{A slice classification neural network for automated classification of axial PET/CT slices from a multi-centric lymphoma dataset
}

\author[1,2,3]{Shadab Ahamed}
\author[3]{Yixi Xu}
\author[5]{Ingrid Bloise}
\author[4]{Joo H. O}
\author[2,5,6]{Carlos F. Uribe}
\author[3]{Rahul Dodhia}
\author[3]{Juan L. Ferres}
\author[1,2,6]{Arman Rahmim}

\affil[1]{Department of Physics \& Astronomy, University of British Columbia, Vancouver, BC, Canada}
\affil[2]{Department of Integrative Oncology, BC Cancer Research Institute, Vancouver, BC, Canada}
\affil[3]{AI for Good Lab, Microsoft, Redmond, WA, USA}
\affil[4]{Seoul St. Mary’s Hospital, The Catholic University of Korea, Seoul, Republic of Korea}
\affil[5]{BC Cancer, Vancouver, BC, Canada}
\affil[6]{Department of Radiology, University of British Columbia, Vancouver, BC, Canada}

\authorinfo{Further author information: (Send correspondence to S. A.)\\S. A.: E-mail: shadabahamed1996@gmail.com}

\begin{document} 
\maketitle
\begin{abstract}
Automated slice classification is clinically relevant since it can be incorporated into medical image segmentation workflows as a preprocessing step that would flag slices with a higher probability of containing tumors, thereby directing physicians’ attention to the important slices. In this work, we train a ResNet-18 network to classify axial slices of lymphoma PET/CT images (collected from two institutions) depending on whether the slice intercepted a tumor (positive slice) in the 3D image or if the slice did not (negative slice). Various instances of the network were trained on 2D axial datasets created in different ways: (i) slice-level split and (ii) patient-level split; inputs of different types were used: (i) only PET slices and (ii) concatenated PET and CT slices; and different training strategies were employed: (i) center-aware (CAW) and (ii) center-agnostic (CAG). Model performances were compared using the area under the receiver operating characteristic curve (AUROC) and the area under the precision-recall curve (AUPRC), and various binary classification metrics. We observe and describe a performance overestimation in the case of slice-level split as compared to the patient-level split training. The model trained using patient-level split data with the network input containing only PET slices in the CAG training regime was the best performing/generalizing model on a majority of metrics. Our models were additionally more closely compared using the sensitivity metric on the positive slices from their respective test sets.
\end{abstract}


\section{INTRODUCTION}
\label{sec:introduction}  
$^{18}$F-Fluorodeoxyglucose positron emission tomography/computed tomography ($^{18}$F-FDG PET/CT) is the current standard of care for imaging lymphoma. Accurate detection and segmentation of lymphoma tumors from PET/CT images have important implications for treatment planning such as radiotherapy and surgical interventions \cite{Slattery2017-ox, Weisman2020-em, sahamed_spie2022}. Tumor segmentation is also required for the quantification of the total metabolic tumor volume (TMTV) - a metric that has predictive value for patient outcomes in lymphoma \cite{Capobianco2021-xs, Cottereau2018-ik, Blanc-Durand2018-fj, Ceriani2018-og}. Manual segmentation of tumors from whole-body PET/CT images is time-consuming and operator-dependent; hence it is not performed routinely \cite{A3346, A3245}. In recent years, the use of deep learning methods has been increasing rapidly in the medical image analysis domain, which can be employed to automate some of these manual tasks \cite{Zhao2018-sv, dl_in_medical_image_analysis, Haenssle2018-pa, Ardila2019-by, Wu2020-br, spie2022_fyr}.\\

Automated slice classification has clinical value since it can be incorporated into medical image segmentation workflows as a preprocessing step that would flag the slices with a higher probability of containing tumors, thereby directing the physician’s attention to the important slices. Our previous work \cite{sahamed_spie2022} performed a classification of axial PET slices into slices containing tumors (positive slices) vs. slices not containing any tumor (negative slices) as a preprocessing step for downstream tasks of tumor localization and segmentation. It demonstrated that an a priori rejection of negative slices was crucial for training the tumor detection and segmentation models since then these modules can be trained on only the positive slices. This made it easier for the detection and segmentation modules to learn contextual representations from the positive slices without being distracted by the noise brought in by the negative examples in the training process.\\

In this work, we implemented, trained, and validated an automated slice classification network. In contrast to the network in our previous work \cite{sahamed_spie2022} that was trained on a smaller uni-centric dataset, this network was trained on a considerably larger and enriched dataset collected at two different institutions (in Canada and South Korea). We trained various models on different types of train-test splits of these datasets, compared their performances using two metrics (i) area under the receiver operating characteristic curve (AUROC), and (ii) area under the precision-recall curve (AUPRC), and discussed ways to improve model generalization, especially to the external cohort. In the end, we evaluate the binary classification metrics for the models based on a specific training methodology and discuss their ability to correctly classify the positive slices.

\section{METHODS}
\label{sec:methods}
\subsection{Dataset}
\label{subsec:dataset}
The lymphoma PET/CT dataset consisted of images collected at two institutions: BC Cancer, Vancouver, Canada (BCCV) ($n=246$; mean age, 49 years; age range, 19-89 years; 129 females), and St. Mary’s Hospital, Seoul, South Korea (SMHS) ($n=220$; mean age, 59 years; age range, 14-87 years; 103 females)\footnote{Both these datasets are not publicly available and would require BC Cancer Research 
Ethics Board (UBC BC Cancer REB) approval for access.}. The lymphoma tumors in both these datasets were collectively annotated by 4 nuclear medicine (NM) physicians at BCCV and 1 NM physician at SMHS. The physicians manually annotated each of the tumor volumes on the PET images using the PETEdge tool from  MIM (MIM Software, Cleveland, Ohio). Note that out of the $n=246$ images from BCCV, only 10 images were collectively annotated by 4 physicians, while the rest were annotated by only one physician (who was one among the four). For those 10 cases, the final ground truth was generated from the individual physician's annotations using the Simultaneous Truth and Performance Level Estimation (STAPLE) algorithm. \cite{staple} \\

The class imbalance between positive and negative axial slices was 8:92 \% and 21:79 \% in the BCCV and SMHS sets, respectively. The higher proportion of positive slices in the SMHS set is due to a larger average number of 3D tumors per patient ($N$) (mean $N$ = 11, range=[1, 128]) and TMTV (mean TMTV = 488.43 ml, range = [1.09, 5919.45] ml), as compared to the BCCV set (mean $N$ = 3, range = [1, 82]) and mean TMTV = 119.25 ml, range = [0.26, 1416.26] ml). A detailed description of the number of axial slices used in various experiments is given in Table \ref{tab:tab2ddata}\\

\subsection{Image preprocessing}
\label{subsec:imagepreprocessing}
Our models were trained on the axial slices of 3D whole-body images. The 3D PET images (attenuation-corrected) were converted into units of SUV$_\text{bw}$ and the CT images were in Hounsfield Units (HU). All the CT images were resampled to have the same size and voxel spacing as PET images. The CT images were also denoised via median filtering with a window size of $5 \times 5 \times 5$ that removed the CT intensity outliers (very large HU values $ \gtrapprox 2000$) while preserving the edges. The proposed network (see next subsection) required a 3-channel input image of size $224 \times 224$. All the PET and CT slices were resized to $224 \times 224$ pixels. The inputs were created to have 3-channels because that was a requirement imposed by the network (as it contained fully-connected layers). Two types of inputs were created for our various experiments in this work: (1) Using only PET slices: input = [\textit{PET slice}, \textit{PET slice}, \textit{PET slice}], and (2) Using both PET and CT slices: input = [\textit{PET slice}, \textit{PET slice}, \textit{CT slice}]. Hence, each input was created by stacking the same axial 2D 1-channel PET slice thrice (as in (1)), or by stacking the same PET slice twice and using the corresponding CT slice as the third channel (as in (2)). The maximum PET intensity value of the tumors in the two datasets was SUV$_\text{bw, max tumor} \approx 48$, so any PET pixel with SUV$_\text{bw} > 50$ was set to $50$, to improve the signal coming from the tumors in the presence of other high(er)-uptake regions such as the brain or bladder. Similarly, the intensities of CT slices were clipped between $[-1024, 1024]$ HU to remove the contribution from bones, calcium, and metal (these have HU $ > 1000$ \cite{LEV2002427}). In the next step, the PET and CT slices were normalized to values between $(0,1)$ using \textit{PET$_\text{norm}$ = PET/50} and \textit{CT$_\text{norm}$ = (CT + 1024)/2048}, respectively before inputting them into the network. Each input slice was labeled as ‘0’ (negative slice) or ‘1’ (positive slice) with the help of the physicians' annotations. \\

\begin{figure} [h]
\centering
\includegraphics[width=0.85\textwidth]{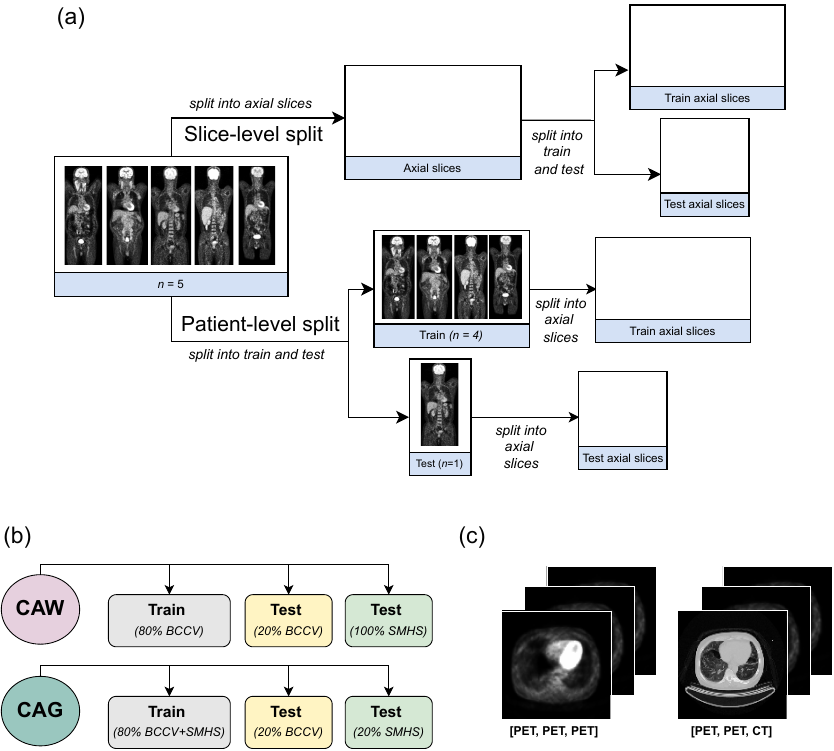}
\vspace{0.3cm}
\caption{(a) Schematic of the slice-level split and patient-level split data, demonstrated using $n=5$ patients. As can be seen in the figure, in the slice-level split, all the 3D images in the dataset are first split into axial slices, and then split into train and test sets. As a result, the train and test sets can have slices from the same patient, leading to the two sets not being independent. On the other hand, the final train and test sets in the patient-level split are always independent. (b) The two training strategies, CAW and CAG. In the CAW strategy, the model is trained only on 80\% BCCV images, while in the CAG strategy, the model is trained on 80\% of the combined dataset from the two centers (BCCV+SMHS). (c) The two types of inputs for network training: using only PET slices (left) with 3-channels formed by concatenating the same PET slice three times and using both PET and CT slices (right) where the first two channels are formed by the repeated PET slice, while the third channel comes from the corresponding CT slices.}
\label{fig:drawing}
\end{figure}

\subsection{Data splitting for model generalization}
\label{subsec:datasplitmodelgeneralization}
The models were trained on two different types of train-test splits of the data. These are explained as follows:
\begin{enumerate}
\item \textbf{Slice-level split}: The 3D images were first split into individual axial slices and then these slices were split 80:20 \% into train and test sets, respectively. Models trained on this kind of data are denoted as $M^S$.
\item \textbf{Patient-level split}: The 3D images were first split 80:20 \% into train and test sets, and then were split into individual axial slices. Models trained on this kind of data are denoted as $M^P$.
\end{enumerate}
For each of the two cases above, two types of models were trained and tested:
\begin{enumerate}
\item \textbf{Center-aware (CAW) training}: In this case, the model was trained on 80\% of BCCV data, and tested on the internal and external test sets, namely the remaining 20\% of BCCV data and 100\% of SMHS data (external cohort). Models trained in this way are denoted as $M_{CAW}$.
\item \textbf{Center-agnostic (CAG) training}: The model was trained on 80\% of combined BCCV+SMHS data and tested on the remaining 20\% of BCCV and 20\% of SMHS data. Models trained in this way are denoted as $M_{CAG}$.    
\end{enumerate}
During training, the training data in each case was further split into training and validation sets (80:20 \%), and the model with the lowest loss on the validation set was chosen as the optimal model for testing.
\begin{table}[h]
\centering
\resizebox{\columnwidth}{!}{%
\begin{tabular}{lccc}
\hline
\multicolumn{1}{|c|}{\textbf{\begin{tabular}[c]{@{}c@{}}Split-type\\ (CAW)\end{tabular}}} &
  \multicolumn{1}{c|}{\textbf{\begin{tabular}[c]{@{}c@{}}Training + Validation set \\ (80\% BCCV)\end{tabular}}} &
  \multicolumn{1}{c|}{\textbf{\begin{tabular}[c]{@{}c@{}}Internal cohort test set\\ (20\% BCCV)\end{tabular}}} &
  \multicolumn{1}{c|}{\textbf{\begin{tabular}[c]{@{}c@{}}External cohort test set \\ (100\% SMHS)\end{tabular}}} \\ \hline
\multicolumn{1}{|l|}{Slice-level split}   & \multicolumn{1}{c|}{50067}  & \multicolumn{1}{c|}{12515} & \multicolumn{1}{c|}{79530} \\ \hline
\multicolumn{1}{|l|}{Patient-level split} & \multicolumn{1}{c|}{50479}  & \multicolumn{1}{c|}{12103} & \multicolumn{1}{c|}{79530} \\ \hline
                                          & \multicolumn{1}{l}{}        & \multicolumn{1}{l}{}       & \multicolumn{1}{l}{}       \\ \hline
\multicolumn{1}{|c|}{\textbf{\begin{tabular}[c]{@{}c@{}}Split-type\\ (CAG)\end{tabular}}} &
  \multicolumn{1}{c|}{\textbf{\begin{tabular}[c]{@{}c@{}}Training + Validation set \\ (80\% BCCV+SMHS)\end{tabular}}} &
  \multicolumn{1}{c|}{\textbf{\begin{tabular}[c]{@{}c@{}}Test set\\ (20\% BCCV)\end{tabular}}} &
  \multicolumn{1}{c|}{\textbf{\begin{tabular}[c]{@{}c@{}}Test set\\ (20\% SMHS)\end{tabular}}} \\ \hline
\multicolumn{1}{|l|}{Slice-level split}   & \multicolumn{1}{c|}{113692} & \multicolumn{1}{c|}{12515} & \multicolumn{1}{c|}{15905} \\ \hline
\multicolumn{1}{|l|}{Patient-level split} & \multicolumn{1}{c|}{114503} & \multicolumn{1}{c|}{12103} & \multicolumn{1}{c|}{15506} \\ \hline
\end{tabular}%
}
\vspace{0.3cm}
\caption{Table showing a description of the number of axial slices in Training + Validation and test sets for slice-level and patient-level split for the CAW (above) and CAG (below) training strategies. During training, the Training + Validation was further split into training and validation sets (80:20 \%).}
\label{tab:tab2ddata}
\end{table}

\begin{figure} [h]
\centering
\includegraphics[width=\textwidth]{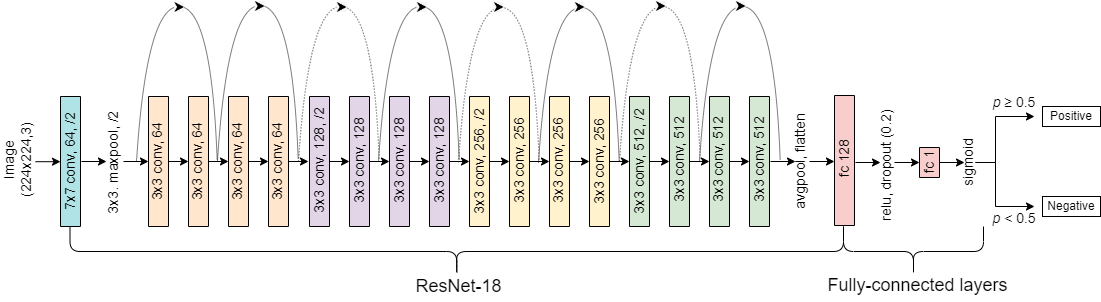}
\caption{Proposed binary classification network with an ImageNet-pretrained ResNet-18 \cite{resnet} module followed by fully-connected layers.}
\label{fig:networkarchitecture}
\end{figure} 

\subsection{Network architecture and training}
\label{subsec:networkarchitecturetraining}
We implemented a binary classification network with an ImageNet-pretrained ResNet-18 \cite{resnet} backbone followed by fully-connected layers ending in a sigmoid, producing an output $p \in (0,1)$ (see Figure \ref{fig:networkarchitecture}). A slice was classified as a positive slice if $p \geq 0.5$, or as a negative slice if $p < 0.5$. The model’s weights were optimized using the Adam optimizer with a constant learning rate=$10^{-3}$. The gradients of the weights of the pretrained ResNet-18 were not frozen and the entire network was fine-tuned during training. Focal loss with hyperparameters $\alpha = 0.25$ and $\gamma = 2$ was used to enhance the binary classification performance in the presence of extreme class imbalance between the positive and negative slices \cite{focalloss}.

\section{RESULTS}
\label{sec:results}
\subsection{Model performance}
\label{subsec:modelperformance}
The performance of the various trained models was compared based on the AUROC and AUPRC values obtained on their respective test sets as shown in Table \ref{tab:tab}. While we report both AUROC and AUPRC in our results, we emphasize that for our datasets with a large class imbalance between the positive and negative slices, the AUPRC metric is more important than AUROC (it is more appropriate to draw performance-based conclusions from the ROC curve when the classes are balanced and from the PRC when the classes are imbalanced). \\

Several interesting conclusions can be drawn from the metrics shown in the table above. Firstly, let’s compare the two cases where the input = [\textit{PET slice}, \textit{PET slice}, \textit{PET slice}]. The CAW model trained on slice-level split data ($M^S_{CAW}$ (20\% BCCV): AUROC=0.99, AUPRC=0.96) outperformed the corresponding CAW model trained on patient-level split data ($M^P_{CAW}$ (20\% BCCV): AUROC=0.89, AUPRC=0.70) on the 20\% BCCV test set (internal test set) by a large margin. This drop in performance can be attributed to the manner the dataset was split into train-test sets in the two cases. For the slice-level split case, different slices from the same patient can exist in training and test sets, making the model memorize the contextual information from a patient in the training set and predict the class correctly with a high probability whenever another nearby slice from the same patient happens to be in the test set, leading to an overestimation of the classification performance. \\

\begin{table}[h]
\centering
\resizebox{\columnwidth}{!}{%
\begin{tabular}{|cccccc|cccccccccccc|}
\hline
\multicolumn{6}{|c|}{\textbf{Slice-level split}} &
  \multicolumn{12}{c|}{\textbf{Patient-level split}} \\ \hline
\multicolumn{6}{|c|}{\textbf{Input = {[}PET, PET, PET{]}}} &
  \multicolumn{6}{c|}{\textbf{Input = {[}PET, PET, PET{]}}} &
  \multicolumn{6}{c|}{\textbf{Input = {[}PET, PET, CT{]}}} \\ \hline
\multicolumn{3}{|c|}{$M^S_{CAW}$} &
  \multicolumn{3}{c|}{$M^S_{CAG}$} &
  \multicolumn{3}{c|}{$M^P_{CAW}$} &
  \multicolumn{3}{c|}{$M^P_{CAG}$} &
  \multicolumn{3}{c|}{$M^P_{CAW}$} &
  \multicolumn{3}{c|}{$M^P_{CAG}$} \\ \hline
\multicolumn{1}{|c|}{\textbf{\begin{tabular}[c]{@{}c@{}}Test\\ set\end{tabular}}} &
  \multicolumn{1}{c|}{\textbf{\begin{tabular}[c]{@{}c@{}}AU\\ ROC\end{tabular}}} &
  \multicolumn{1}{c|}{\textbf{\begin{tabular}[c]{@{}c@{}}AU\\ PRC\end{tabular}}} &
  \multicolumn{1}{c|}{\textbf{\begin{tabular}[c]{@{}c@{}}Test\\  set\end{tabular}}} &
  \multicolumn{1}{c|}{\textbf{\begin{tabular}[c]{@{}c@{}}AU\\ ROC\end{tabular}}} &
  \textbf{\begin{tabular}[c]{@{}c@{}}AU\\ PRC\end{tabular}} &
  \multicolumn{1}{c|}{\textbf{\begin{tabular}[c]{@{}c@{}}Test\\  set\end{tabular}}} &
  \multicolumn{1}{c|}{\textbf{\begin{tabular}[c]{@{}c@{}}AU\\ ROC\end{tabular}}} &
  \multicolumn{1}{c|}{\textbf{\begin{tabular}[c]{@{}c@{}}AU\\ PRC\end{tabular}}} &
  \multicolumn{1}{c|}{\textbf{\begin{tabular}[c]{@{}c@{}}Test\\  set\end{tabular}}} &
  \multicolumn{1}{c|}{\textbf{\begin{tabular}[c]{@{}c@{}}AU\\ ROC\end{tabular}}} &
  \multicolumn{1}{c|}{\textbf{\begin{tabular}[c]{@{}c@{}}AU\\ PRC\end{tabular}}} &
  \multicolumn{1}{c|}{\textbf{\begin{tabular}[c]{@{}c@{}}Test\\  set\end{tabular}}} &
  \multicolumn{1}{c|}{\textbf{\begin{tabular}[c]{@{}c@{}}AU\\ ROC\end{tabular}}} &
  \multicolumn{1}{c|}{\textbf{\begin{tabular}[c]{@{}c@{}}AU\\ PRC\end{tabular}}} &
  \multicolumn{1}{c|}{\textbf{\begin{tabular}[c]{@{}c@{}}Test\\  set\end{tabular}}} &
  \multicolumn{1}{c|}{\textbf{\begin{tabular}[c]{@{}c@{}}AU\\ ROC\end{tabular}}} &
  \textbf{\begin{tabular}[c]{@{}c@{}}AU\\ PRC\end{tabular}} \\ \hline
\multicolumn{1}{|c|}{\textbf{\begin{tabular}[c]{@{}c@{}}20\% \\ BCCV\end{tabular}}} &
  \multicolumn{1}{c|}{0.99} &
  \multicolumn{1}{c|}{0.96} &
  \multicolumn{1}{c|}{\textbf{\begin{tabular}[c]{@{}c@{}}20\% \\ BCCV\end{tabular}}} &
  \multicolumn{1}{c|}{0.99} &
  0.97 &
  \multicolumn{1}{c|}{\textbf{\begin{tabular}[c]{@{}c@{}}20\% \\ BCCV\end{tabular}}} &
  \multicolumn{1}{c|}{0.89} &
  \multicolumn{1}{c|}{0.70} &
  \multicolumn{1}{c|}{\textbf{\begin{tabular}[c]{@{}c@{}}20\% \\ BCCV\end{tabular}}} &
  \multicolumn{1}{c|}{0.91} &
  \multicolumn{1}{c|}{0.71} &
  \multicolumn{1}{c|}{\textbf{\begin{tabular}[c]{@{}c@{}}20\% \\ BCCV\end{tabular}}} &
  \multicolumn{1}{c|}{0.90} &
  \multicolumn{1}{c|}{0.70} &
  \multicolumn{1}{c|}{\textbf{\begin{tabular}[c]{@{}c@{}}20\% \\ BCCV\end{tabular}}} &
  \multicolumn{1}{c|}{0.91} &
  0.71 \\ \hline
\multicolumn{1}{|c|}{\textbf{\begin{tabular}[c]{@{}c@{}}100\% \\ SMHS\end{tabular}}} &
  \multicolumn{1}{c|}{0.81} &
  \multicolumn{1}{c|}{0.63} &
  \multicolumn{1}{c|}{\textbf{\begin{tabular}[c]{@{}c@{}}20\% \\ SMHS\end{tabular}}} &
  \multicolumn{1}{c|}{0.99} &
  0.98 &
  \multicolumn{1}{c|}{\textbf{\begin{tabular}[c]{@{}c@{}}100\% \\ SMHS\end{tabular}}} &
  \multicolumn{1}{c|}{0.86} &
  \multicolumn{1}{c|}{0.71} &
  \multicolumn{1}{c|}{\textbf{\begin{tabular}[c]{@{}c@{}}20\% \\ SMHS\end{tabular}}} &
  \multicolumn{1}{c|}{0.92} &
  \multicolumn{1}{c|}{0.85} &
  \multicolumn{1}{c|}{\textbf{\begin{tabular}[c]{@{}c@{}}100\% \\ SMHS\end{tabular}}} &
  \multicolumn{1}{c|}{0.83} &
  \multicolumn{1}{c|}{0.68} &
  \multicolumn{1}{c|}{\textbf{\begin{tabular}[c]{@{}c@{}}20\% \\ SMHS\end{tabular}}} &
  \multicolumn{1}{c|}{0.91} &
  0.86 \\ \hline
\end{tabular}%
}
\vspace{0.3cm}
\caption{Table showing the performance of various trained classification models on their various test sets using AUROC and AUPRC metrics. The models $M^S_{CAW}$ and $M^P_{CAW}$ have been tested on 20\% BCCV (internal cohort) and 100\% SMHS (external cohort), while the models $M^S_{CAG}$ and $M^P_{CAG}$ have been test on 20\% BCCV and 20\% SMHS test sets. On the slice-level data, the models overestimate the performance since the training and test datasets are not independent. The performance of the $M_{CAG}$ models was always better than their respective $M_{CAW}$ models showing that the datasets from the two institutions were quite different and it was important to include slices from SMHS to the training set to make the models generalize better. We also noted that using CT information provided practically no gain in classification performance (and also deteriorated the performance by a few percentages on some of the metrics.} 
\label{tab:tab}
\vspace{0.4cm}
\end{table}

\begin{figure}[h]
\centering
\includegraphics[width=0.8\textwidth]{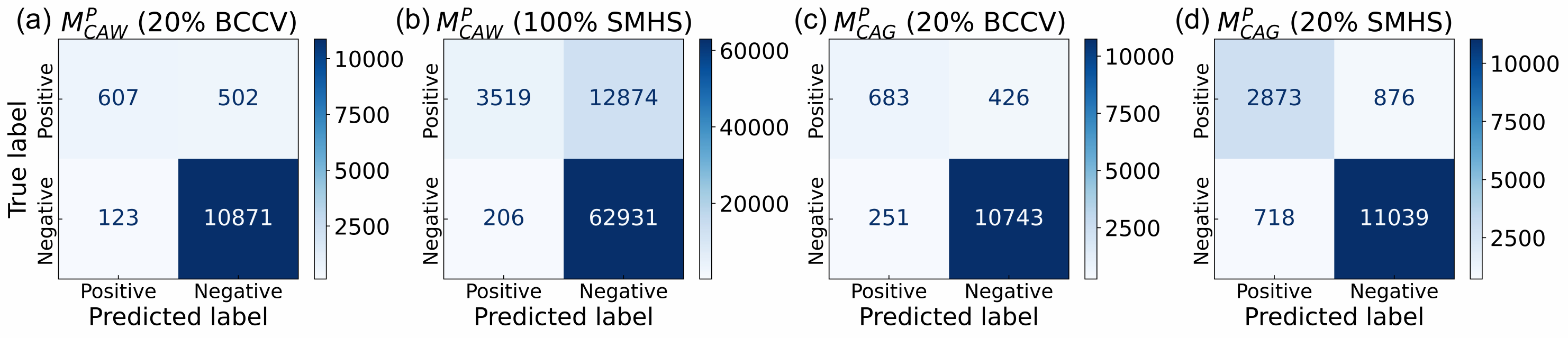}
\caption{Confusion matrices for the models (a) $M^P_{CAW}$ (20\% BCCV), (b) $M^P_{CAW}$ (100\% SMHS), (c) $M^P_{CAG}$ (20\% BCCV), and (d) $M^P_{CAG}$ (20\% SMHS), evaluated on their respective test sets (given in brackets).}
\label{fig:confusionmatrices}
\end{figure} 

On the other hand, in the patient-level split case, the training and test sets are totally independent of each other at the patient level. A similar conclusion can be drawn when comparing the CAG models where the models trained on slice-level split data (($M^S_{CAG}$ (20\% BCCV): AUROC=0.99, AUPRC=0.97) and ($M^S_{CAG}$ (20\% SMHS): AUROC=0.99, AUPRC=0.98)) outperformed those trained on the patient-level split data (($M^P_{CAG}$ (20\% BCCV): AUROC=0.91, AUPRC=0.71) and ($M^P_{CAG}$ (20\% SMHS): AUROC=0.92, AUPRC=0.85)). Hence, our results suggest that when training 2D models, one must perform a patient-level split of their data to prevent performance overestimation.\\

It is also interesting to note that despite this drop in performance for the CAW model on the internal test cohort when going from slice-level split to patient-level split data, the CAW model trained on patient-level split data ($M^P_{CAW}$ (100\% SMHS): AUROC=0.86, AUPRC=0.71) generalized better than the CAW model trained on slice-level split data ($M^S_{CAW}$ (100\% SMHS): AUROC=0.81, AUPRC=0.63) on the 100\% SMHS test set (external cohort). \\

Secondly, we also trained models on the patient-level split data with input = [\textit{PET slice}, \textit{PET slice}, \textit{CT slice}] so as to utilize the anatomical information from the CT slices. There wasn’t much difference in the performance metrics upon replacing a PET slice with a CT slice as one of the channels, except for $M^P_{CAW}$ (100\% SMHS) where there was a drop of 3\% in both AUROC and AUPRC for external validation upon including CT. These findings can be explained in one of the following ways: (a) the replacement of the PET slice with a CT slice in the 3-channel image reduced the overall signal in the image from the tumor/background on the positive/negative slices; (b) the clipping of CT slice intensities in the range [-1024, 1024] HU might not be an optimal preprocessing for this problem, and other preprocessing methods must be investigated; (c) the current dataset is not large enough to efficiently utilize the anatomical information from the CT images.\\

It can also be concluded that the models trained on just the 80\% BCCV data ($M^S_{CAW}$ and $M^P_{CAW}$) do not generalize well to the SMHS data, and hence including SMHS data in the training process is crucial for model generalization as it can be seen from the higher classification performances of $M^S_{CAG}$ and $M^P_{CAG}$ on 20\% SMHS data as compared to the that of $M^S_{CAW}$ and $M^P_{CAW}$ on 100\% SMHS data (external test cohort).\\

\begin{figure}[h]
\vspace{0.6cm}
\centering
\includegraphics[width=0.9\textwidth]{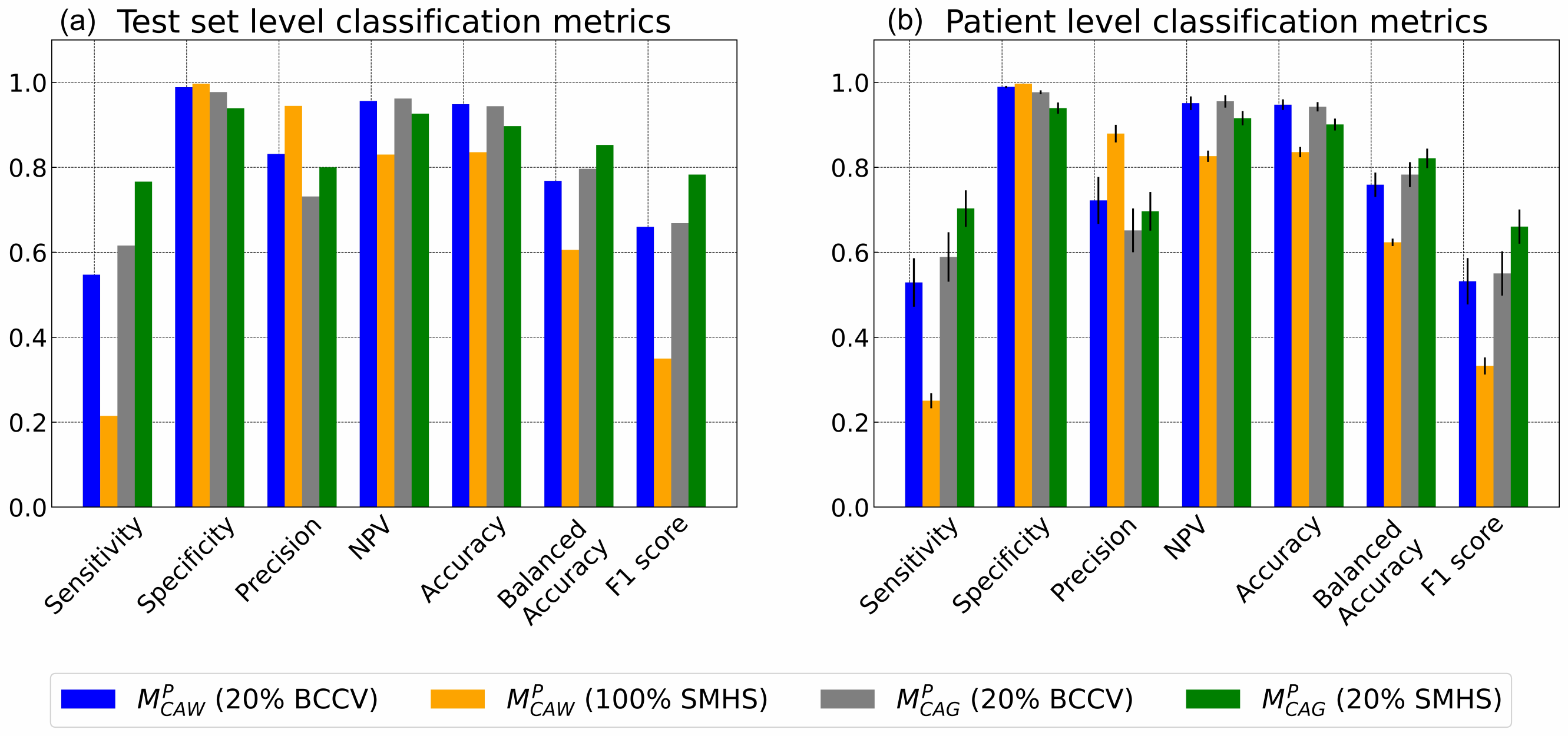}
\caption{Evaluation of trained binary classifiers $M^P_{CAW}$ (20\% BCCV), $M^P_{CAW}$ (100\% SMHS), $M^P_{CAG}$ (20\% BCCV), and $M^P_{CAG}$ (20\% SMHS) on their respective test sets (given in brackets). Classification metrics were evaluated at the test set level (a) and at the patient level (b) within these test sets.}
\label{fig:classificationmetrics}
\end{figure} 

\subsection{Classification metrics}
\label{subsec:classificationmetrics}
The models trained on patient-level split data ($M^P_{CAW}$ and $M^P_{CAG}$) with inputs as [\textit{PET slice}, \textit{PET slice}, \textit{PET slice}] were evaluated with binary classification metrics such as sensitivity, specificity, precision, negative predictive value (NPV), accuracy, balanced accuracy, and F1 score. The confusion matrices for the cases $M^P_{CAW}$ (20\% BCCV), $M^P_{CAW}$ (100\% SMHS), $M^P_{CAG}$ (20\% BCCV), and $M^P_{CAG}$ (20\% SMHS) are given in Figures \ref{fig:confusionmatrices}(a), \ref{fig:confusionmatrices}(b), \ref{fig:confusionmatrices}(c), and \ref{fig:confusionmatrices}(d), respectively. The classification metrics were computed at the test set level in Figure \ref{fig:classificationmetrics}(a) and at the patient level within the test sets in Figure \ref{fig:classificationmetrics}(b). It can be seen that the performances of $M^P_{CAW}$ (20\% BCCV) and $M^P_{CAG}$ (20\% BCCV) are nearly the same for all metrics, while the $M^P_{CAG}$ (20\% SMHS) in general outperforms $M^P_{CAW}$ (100\% SMHS) on most metrics (except for specificity and precision). This is because $M^P_{CAW}$ was trained only on the 80\% BCCV set, it failed to learn representative features of the SHMS set, while $M^P_{CAG}$ being trained on a combination of BCCV and SMHS sets was good at predicting the correct class for the slices from both the institutions. This could be attributed to a number of large variabilities between the dataset from the two centers such as differences in average TMTV and the average number of tumors per patient. Interestingly, our network architecture and training methodology leads to a very low number of false positive predictions, resulting in considerably high specificity values ($>0.95$) at both the test set and patient level for all the models.  \\

\begin{figure}[H]
\centering
\includegraphics[width=0.7\textwidth]{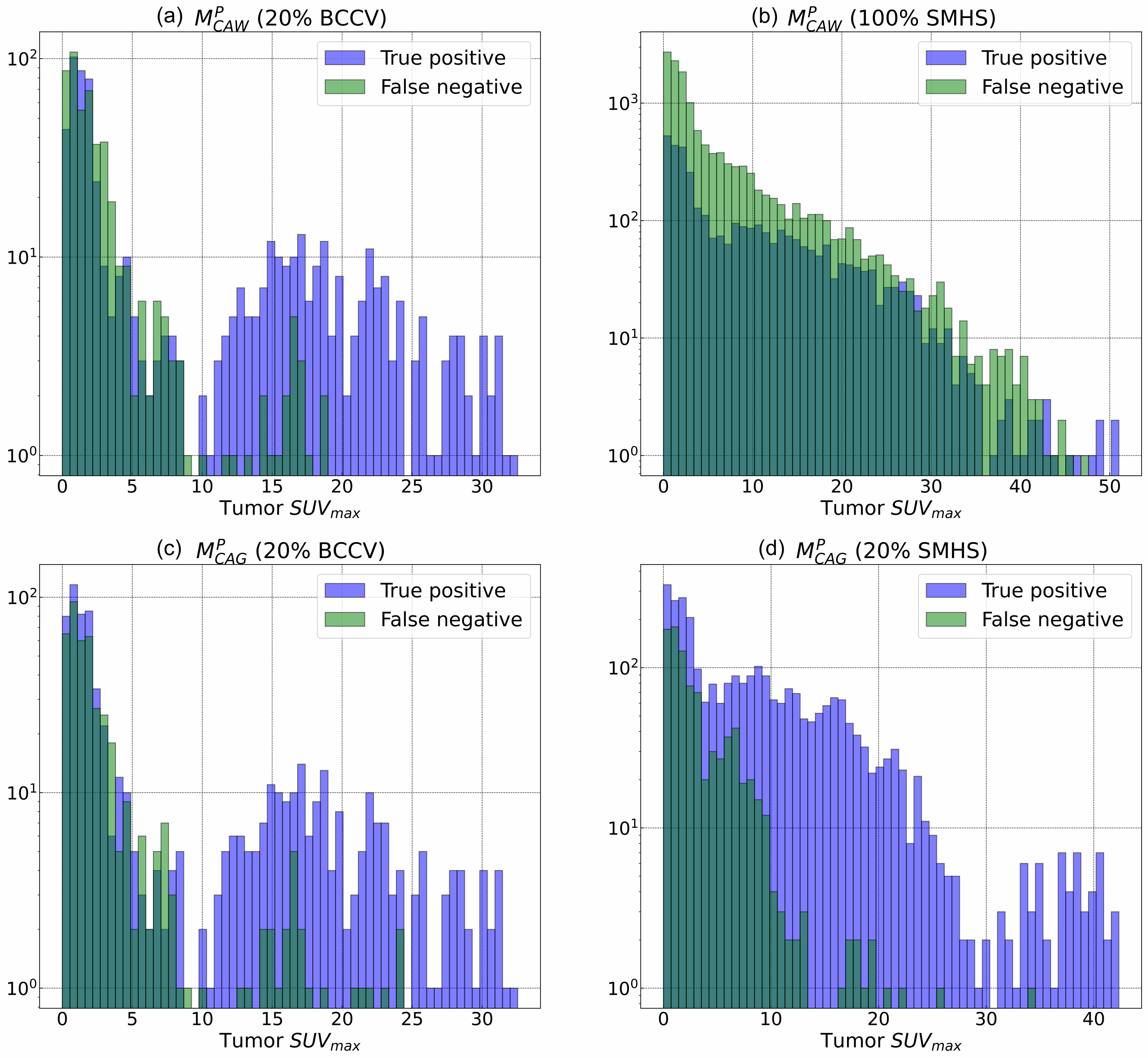}
\caption{The histograms show the distribution of the number of true positive and false negative predictions as a function of tumor SUV$_\text{max}$ for (a) $M^P_{CAW}$ (20\% BCCV), (b) $M^P_{CAW}$ (100\% SMHS), (c) $M^P_{CAG}$ (20\% BCCV), and (d) $M^P_{CAG}$ (20\% SMHS) on their respective test sets (given in brackets).}
\label{fig:fgslicehistograms}
\end{figure}

\subsection{Assessing the classification performance for positive slices (sensitivity analysis)}
\label{subsec:sensitivityanalysis}
The classification performance was more closely investigated for the positive slices. The SUV$_\text{max}$ values of the tumors on the positive slices were noted and its dependence on the model’s classification performance (into true positive or false negative) was evaluated for the cases $M^P_{CAW}$ (20\% BCCV), $M^P_{CAW}$ (100\% SMHS), $M^P_{CAG}$ (20\% BCCV), and $M^P_{CAG}$ (20\% SMHS) as shown in the histograms in Figures \ref{fig:fgslicehistograms}(a), \ref{fig:fgslicehistograms}(b), \ref{fig:fgslicehistograms}(c), and \ref{fig:fgslicehistograms}(d), respectively. The histograms for $M^P_{CAW}$ (20\% BCCV), $M^P_{CAG}$ (20\% BCCV), and $M^P_{CAG}$ (20\% SMHS) show that a larger number of positive slices were classified correctly that had larger values of tumor SUV$_\text{max}$, while there was a similar proportion of true positives and false negatives for slices with lower SUV$_\text{max}$. We hypothesize that the correct classification of some of the slices with low SUV$_\text{max}$ is due to the fact that these slices intercepted a larger portion of the 3D tumor (ongoing efforts). On the contrary, the histogram for $M^P_{CAW}$ (100\% SMHS) shows that in this case most of the positive slices were classified as false negatives, which is also the reason for the considerably low sensitivity for $M^P_{CAW}$ (100\% SMHS) as compared to $M^P_{CAW}$ (20\% BCCV), $M^P_{CAG}$ (20\% BCCV), and $M^P_{CAG}$ (20\% SMHS).\\

\begin{figure}[H]
\centering
\includegraphics[width=0.7\textwidth]{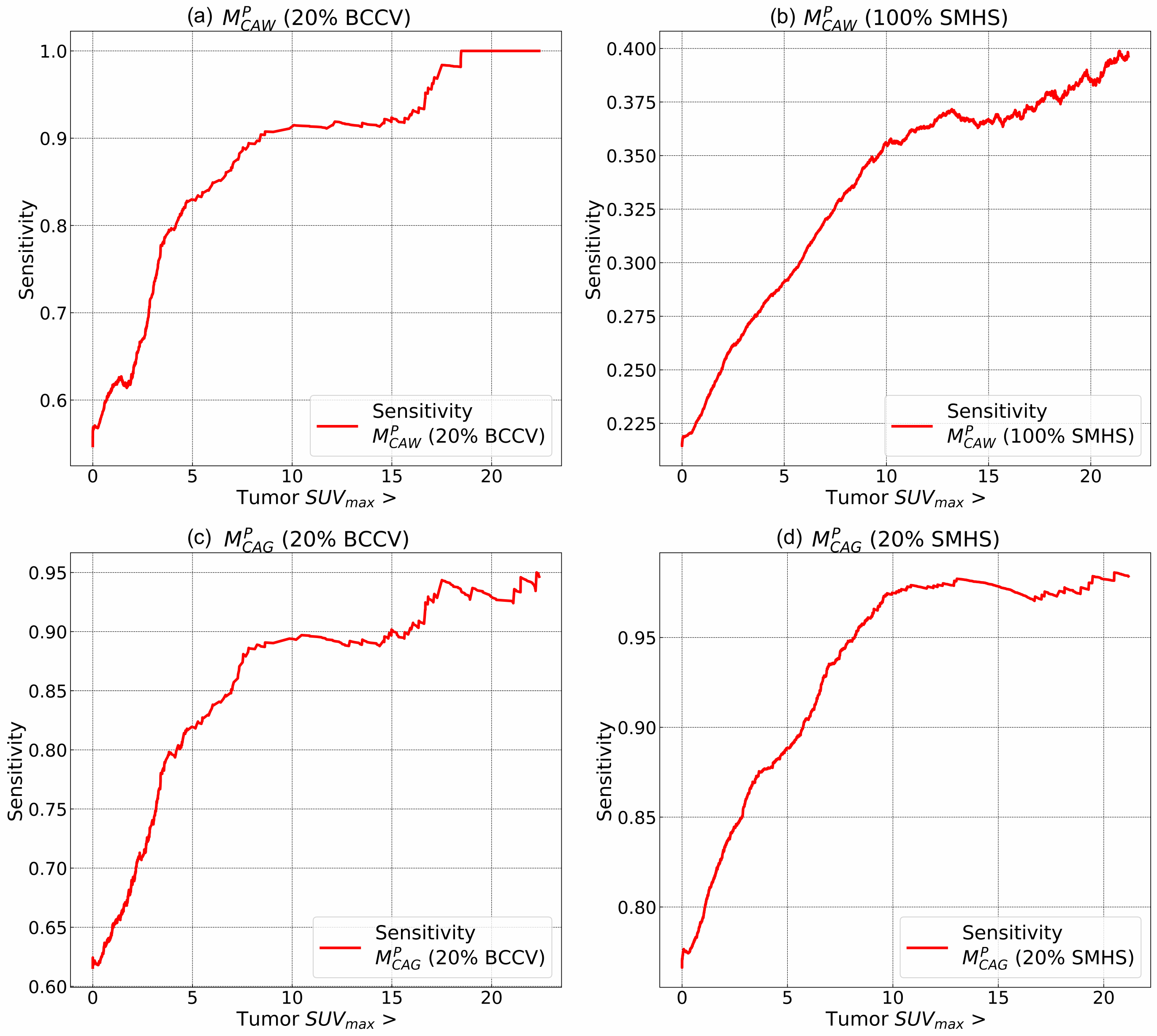}
\caption{Sensitivity or true positive rate analysis as a function of tumor SUV$_\text{max}$ for the models (a) $M^P_{CAW}$ (20\% BCCV), (b) $M^P_{CAW}$ (100\% SMHS), (c) $M^P_{CAG}$ (20\% BCCV), and (d) $M^P_{CAG}$ (20\% SMHS) on their respective test sets (shown in brackets). Here, we unequivocally demonstrate that, as expected, there is an increase in sensitivity as a function of tumor SUV$_\text{max} > u_i$, where $u_i$ is the minimum tumor SUV$_\text{max}$ value among all the slices within the smaller test subset $i$ (where $1 \leq i \leq \lfloor 0.95m \rfloor$ and $m$ is the total number of slices in the test set).}
\label{fig:fgslicesensitivities}
\end{figure}

Finally, the positive slices in each of these test sets were arranged in increasing order of SUV$_\text{max}$ such that $u_1 \leq u_2 \leq \cdots \leq u_{m-1} \leq u_m$, where $u_i$ is the tumor SUV$_\text{max}$ value of the positive slice with the $i^\text{th}$ largest SUV$_\text{max}$ and $m$ is the total number of slices in the test set. Within each test set, different subsets of slices were chosen such that subset $i$ consisted of all the slices with SUV$_\text{max} \geq u_i$. Hence, the subset $i = 1$ consisted of all $m$ slices, while the subset $i = m$ consisted of only the slice with SUV$_\text{max} = u_m$. The sensitivity or true positive rate (= true positives/(total number of positives)) was computed on each of these subsets and plotted as a function of $u_i$ for $1 \leq i \leq \lfloor 0.95m \rfloor$, for each of $M^P_{CAW}$ (20\% BCCV), $M^P_{CAW}$ (100\% SMHS), $M^P_{CAG}$ (20\% BCCV), and $M^P_{CAG}$ (20\% SMHS), as shown in Figures \ref{fig:fgslicesensitivities}(a), \ref{fig:fgslicesensitivities}(b), \ref{fig:fgslicesensitivities}(c), and \ref{fig:fgslicesensitivities}(d), respectively. Each of these plots unequivocally shows that the percentage of correctly classified positive slices increases with an increase in tumor SUV$_\text{max}$ in the subset, showing that, as expected, the networks found it easier to correctly classify the positive slices containing tumors with higher SUV$_\text{max}$.

\section{DISCUSSION and CONCLUSION}
\label{sec:conclusiondiscussion}
This study may also be extended by doing a similar analysis as in Figures \ref{fig:fgslicesensitivities}(a)-\ref{fig:fgslicesensitivities}(d) for the negative slices, where the SUV$_\text{max}$ for a negative slice can be computed as the maximum non-tumorous SUV value within that slice. Furthermore, other classifiers with different pretrained backbone networks will be explored towards enhancing the classification performance further.\\

We trained different instances of a binary classifier neural network for the various train-test splits of the data and explored the model's generalizability (via two training regimes, CAW and CAG) in the context of a multi-centric PET/CT dataset. We showed that the inclusion of CT slices provided no performance boost, at least for the models we trained with inputs containing a CT slice as the third channel. Further investigation into different CT preprocessing might be required to make the model efficiently use the rich anatomical information provided by the CT images. \\

We demonstrated unequivocally that the tumor SUV$_\text{max}$ on positive slices is important for the deep learning classifier to predict the class for positive slices. We also noted that the positive slices with larger tumor SUV$_\text{max}$ values are more likely to be classified correctly, as compared to the positive slices with lower tumor SUV$_\text{max}$. It is worth noting that the tumor SUV$_\text{max}$ value is highly dependent on the PET reconstruction algorithms used. While this study did not consider the effects of different reconstruction algorithms used for different images from the two centers, this will be taken up as a separate study in future works.\\

\section*{ACKNOWLEDGEMENTS}
\label{sec:acknowledgements}
This work was supported by the Canadian Institutes of Health Research (CIHR) Project Grant PJT-173231, the Mitacs Accelerate grant, and computational resources and services provided by Microsoft for Health.

\bibliography{report} 
\bibliographystyle{spiebib} 

\end{document}